\documentclass[aps,prl,twocolumn,reprint,superscriptaddress]{revtex4}
\usepackage{graphicx}
\usepackage{amsmath}
\usepackage{amssymb}
\usepackage{colordvi}
\usepackage{mathrsfs}
\usepackage{bm}
\usepackage{verbatim}
\usepackage{dcolumn}
\usepackage{epsfig}
\usepackage{subfigure}

\begin{document}
\title{Physical Origin of Current Partition at a Topological Trifurcation}
\author{Sanyi You}
\affiliation{ICQD, Hefei National Laboratory for Physical Sciences at Microscale, CAS Key Laboratory of Strongly-Coupled Quantum Matter Physics, and Department of Physics, University of Science and Technology of China, Hefei, Anhui 230026, China}
\author{Tao Hou}
\affiliation{ICQD, Hefei National Laboratory for Physical Sciences at Microscale, CAS Key Laboratory of Strongly-Coupled Quantum Matter Physics, and Department of Physics, University of Science and Technology of China, Hefei, Anhui 230026, China}
\author{Zhenhua Qiao}
\email[Correspondence author:~]{qiao@ustc.edu.cn}
\affiliation{ICQD, Hefei National Laboratory for Physical Sciences at Microscale, CAS Key Laboratory of Strongly-Coupled Quantum Matter Physics, and Department of Physics, University of Science and Technology of China, Hefei, Anhui 230026, China}
\date{\today{}}

\begin{abstract}
  {In gated bilayer graphene, topological zero-line modes (ZLMs) appear along
  lines separating regions with opposite valley Hall topologies. Although it is
  experimentally difficult to design the electric gates to realize ZLMs due to
  the extremely challenging techniques, twisted bilayer graphene provides a
  natural platform to produce ZLMs in the presence of uniform electric
  field. In this Letter, we develop a set of wavepacket dynamics, which can be
  utilized to characterize various gapless edge modes and can quantitatively
  reproduce the electronic transport properties at topological intersections.
  To our surprise, in the minimally twisted bilayer graphene where a
  topological trifurcation intersection naturally arises, we show that the
  counterintuitive current partition (i.e., the direct transport propagation)
  originates from the microscopic mechanism ``bypass jump". Our method can be
  applied to understand the microscopic pictures of
  the electronic transport features of all kinds of topological states.}
\end{abstract}
\maketitle

\textit{Introduction---.} Topological zero-line modes (ZLMs) can arise at
interfaces separating different topological systems, e.g., classical wave
systems~\cite{cws1,cws2,cws3,cws4,cws5,cws6}, non-Hermitian
systems~\cite{nhs1,nhs2}, and various graphene
systems~\cite{vgs1,vgs2,vgs3,vgs4,vgs5,vgs6,vgs7,vgs8,vgs9,vgs10,vgs11,vgs12,vgs13,vgs14,vgs15,vgs16,vgs17,vgs18,vgs19,vgs20,vgs21,vgs22}.
 Because of their robustness against backscattering, ZLMs have attracted
numerous attention in designing low-power topological quantum devices. In
particular, gated AB-stacked bilayer graphene acts as an ideal platform in
generating the ZLMs along zero-field line separating the regions with opposite
valley Hall topologies. However, the requirements of extremely precise
alignment of the electric gates in bilayer graphene made it unrealistic for
large-scale industrial application. Fortunately, the manipulation of
``twisting" made bilayer graphene again the central focus in both theoretical
and experimental condensed matter physics. The twisted bilayer graphene then
naturally provides an ideal platform (i.e., metallic moir$\acute{\rm e}$
pattern networks) in designing ZLMs in the presence of uniform electric
field~\cite{tbs1,tbs2,tbs3,tbs4,tbs5}.

So far, although there have been great progress in exploring the electronic transport properties of the ZLMs from both numerical calculations and experiments from a macro perspective, it is still analytically unsolvable whenever the zero-line becomes curved or crossing. The reason is that it is extremely challenging to derive the analytical solution of the two-dimensional Dirac equation for curved ZLMs or arbitrary topological ZLM intersections~\cite{vgs18}. Therefore, the fundamental physical understanding of the transport characteristics of ZLMs is still missing.

Time-dependent Schr\"{o}dinger equation and wavepacket dynamics are
time-honored research topics. They provide powerful tools in investigating the
motion and scattering problems of quantum particles from an intuitive and
time-dependent perspective. If one treats the electron as a wavepacket, one can
obtain the position and velocity of wavepacket center at any time. In several
low-dimensional materials, the wavepacket propagations have been
investigated~\cite{wpp1,wpp2,wpp3,wpp4,wpp5}, but only the evolution of a
simple Gaussian wavepacket is adopted in most cases~\cite{wpp2,wpp3,wpp4,wpp5}.
However, the Gaussian wavepacket is not the eigenstate of the Hamiltonian of
topological system, but leads to strong dissipation during the evolution.

\begin{figure}[!htp]
  \includegraphics[width=8.5cm,angle=0]{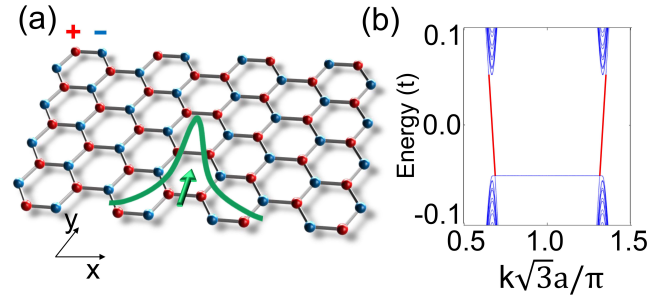}
  \caption{(a) Schematic plot of a single straight zero line in monolayer
  graphene. Red and blue denote ``+/-"
  sublattice  potentials, respectively. (b) Band structures for zero line in
  monolayer zigzag nanoribbon.}
  \label{Fig1}
\end{figure}

In this Letter, we develop a set of wavepacket dynamics for the topological
zero line systems, which can clearly describe the time-dependent evolution of
the ZLM at topological intersections. By tracking the trajectories of the
electron wavepackets at various topological intersections, one can obtain the
corresponding current partition ratios in an intuitive manner. In particular,
in the minimally twisted bilayer graphene system, we find that the incoming
wavepacket is divided into three parts at the intersection, where the direct
transmission tunneling is not so ``direct" along AA-stacked regions but
originates from the ``bypass jump" scattering mechanism. Our proposed methods
provide a powerful tool in revealing the time-dependent trajectories of
electrons and judging the transport properties in topological materials.

\begin{figure}[!htp]
  \includegraphics[width=8.5cm,angle=0]{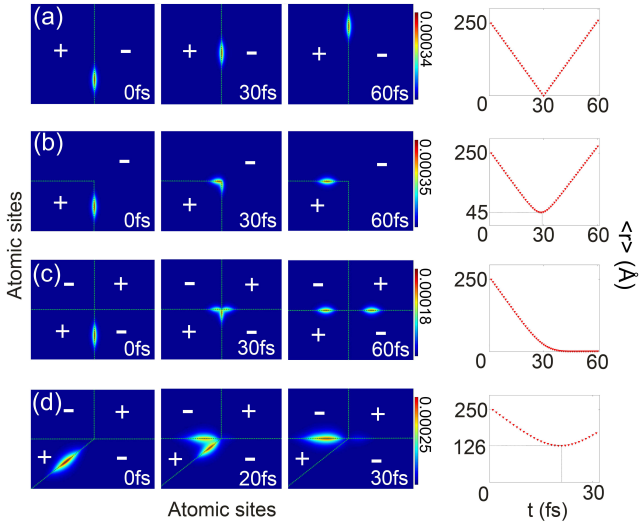}
  \caption{Time evolution of the wavepacket and $\langle r \rangle$ for (a) a
  single straight zero line, (b) a right-angled zero line, (c) crisscross zero
  lines, and (d) a bifurcation with 30$^{\circ}$ sharp angle. The sample sizes
  are $L$ = 170.3 nm and $W$ = 98.4 nm in (a)-(c), and $L$ = 97.0 nm and $W$ =
  56.1 nm in (d). $d_1$ = 4.2 nm, $d_2$ = 10.0 nm. $\langle r_0 \rangle$ = 25.0
  nm. The + and - signs indicate the alternating sublattice potentials.}
  \label{Fig2}
\end{figure}

\begin{figure}[!htp]
  \includegraphics[width=8.5cm,angle=0]{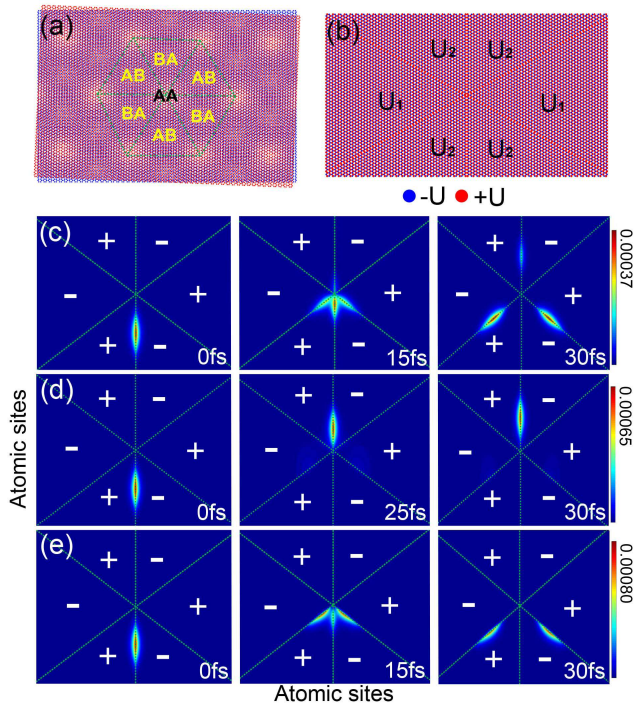}
  \caption{(a) Schematic view of twisted bilayer graphene. The AA (in black)
  rigions denote the topological intersection. The green lines denote the
  topological zero lines that separate alternating AB and BA regions
  distinguished by local valley Chern numbers. (b) Equivalent simulation of
  topological intersection in monolayer graphene with sublattice-staggered
  potentials labeled in blue and red. (c)-(e) Time evolution of the wavepacket.
  We have $U_1=U_2=\lambda t$ in (c), $U_1=0.1U_2=0.1\lambda t$ in (d) and
  $U_1=4U_2=4\lambda t$ in (e), where $\lambda$ =0.08. The sample sizes are $L$
  = 97.4 nm and $W$ = 56.8 nm in (c)-(e). $d_1$ =2.6 nm, $d_2$ = 6.0 nm.
  $\langle r_0 \rangle$ = 14.0 nm. The + and - signs indicate the alternating
  sublattice potentials.}
  \label{Fig3}
\end{figure}

\begin{figure}[!htp]
\includegraphics[width=8.5cm,angle=0]{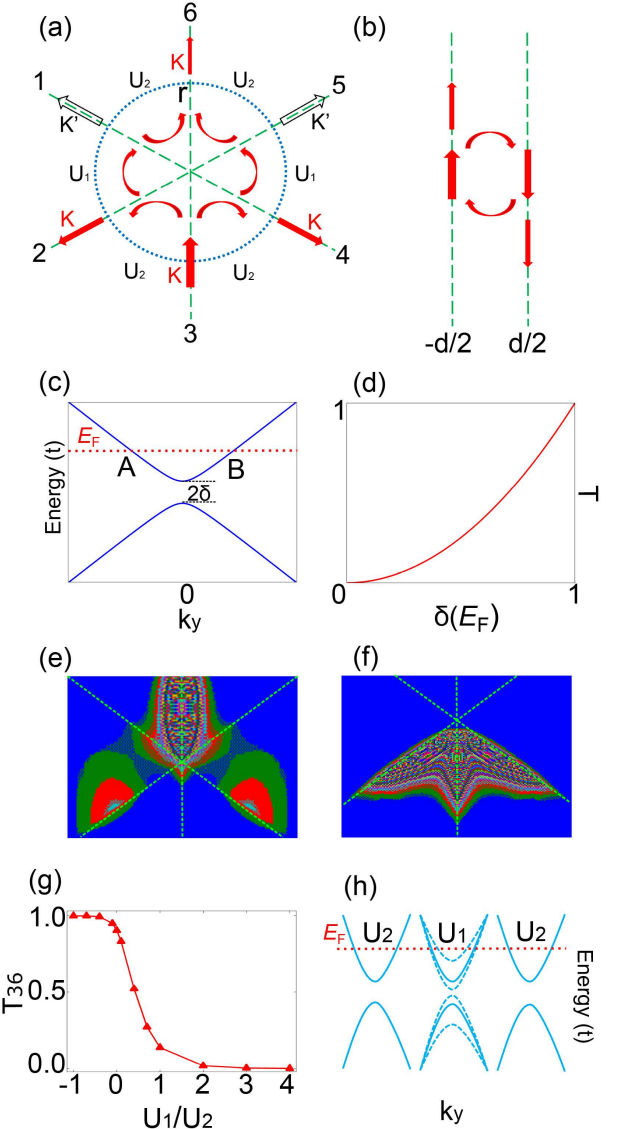}
\caption{(a) Simplified schematic of Fig.~\ref{Fig3}(b), the red and black
arrows represents the permitted and not permitted scattering direction of the
electron
wavepacket. (b) At a
circumference with a given $r$, sketch of the tunneling between adjacent
topological zero lines with a distance of d. (c) Low-energy spectrum near
valley K of two topological zero lines illustrated in panel (b) with a gap
induced by the finite-size effect. The red dashed line indicates the Fermi
level. (d)
Scattering rate versus gap $\delta$. (e)-(f) Zoom of the second subgraph of
Fig.~\ref{Fig3}(d)-Fig.~\ref{Fig3}(e). (g) Scattering rate from terminal 3 to
terminal 6 versus the ratio of $U_1/U_2$. (h) At a circumference with a
given $r$, the evolution of band structure from terminal 3 to terminal 6.}
\label{Fig4}
\end{figure}

\textit{Time-evolution operator---.} To numerically investigate the electronic properties of ZLMs, the $\pi$-orbital tight-binding model Hamiltonian is adopted as following:
\begin{eqnarray}
  H=-t\sum_{<ij>}c_i^\dagger c_j+\sum_{i\in A}U_{A}c_i^\dagger c_i+\sum_{i\in
  B}U_{B}c_i^\dagger c_i
\label{eq1}
\end{eqnarray}
where $c_i^\dagger$($c_i$) is a creation (annihilation) operator for an electron at site $i$, and $t$ = 2.7 eV is the nearest-neighbor hopping energy. $U_{\rm A}$ and $U_{\rm B}$ are the staggered AB sublattice site potentials, satisfying $U_{A}=-U_{B}=\lambda t$. We set $\lambda$ = 0.05t in Fig.~\ref{Fig2}, and 0.08t in Figs.~\ref{Fig3} and \ref{Fig4}.

Based on the time-evolution operator, if the initial wavefunction $\Psi(x,y,t)$ is known, the propagated wavefunction at each time step $\Delta$$t$ becomes
\begin{eqnarray}
\Psi(x,y,t+\Delta t)=\exp(-\frac{i}{\hbar} H {\Delta t})\Psi(x,y,t).
\label{eq2}
\end{eqnarray}
By applying the Cayley form~\cite{cfe}, one can obtain
\begin{eqnarray}
(1+\frac{i}{2\hbar}H\Delta t)\Psi(x,y,t+\Delta t)\approx(1-\frac{i}{2\hbar}H\Delta t)\Psi(x,y,t).
\label{eq3}
\end{eqnarray}

When different zero lines are orthogonal or parallel to each other, it is possible to apply the spilt-operator technique~\cite{wpp2,wpp3}, which can easily tridiagonalise the full Hamiltonian. The full Hamiltonian is usually split into two parts, one for hopping along $x$ direction and the other along $y$ direction, i.e., $H\Psi=H_x\Psi_x+H_y\Psi_y$, where $\Psi$ is the wavefunction of the whole system and $\Psi_{x/y}$ is the wavefunction along $x/y$ direction. Then, the time-evolution operator can be rewritten as
\begin{widetext}
\begin{eqnarray}
\exp{({-i\frac{H}{\hbar} \Delta t})} =\exp({-\frac{i}{2\hbar}H_y\Delta t})
\exp({-\frac{i}{\hbar}H_x\Delta
t})\exp({-\frac{i}{2\hbar}H_y\Delta t})+\emph{O}(\Delta t^{3}).
\label{eq4}
\end{eqnarray}
\end{widetext}

By substituting the above expression into Eq.~(\ref{eq2}), and then repeating the process in Eq.~(\ref{eq3}), $\Psi(x,y,t+\Delta t)$ can be finally obtained.

\textit{Initial Wave Function---.} As a tentative solution, a symmetric Gaussian wavepacket is not suitable for zero-line mode due to the
confined effect in the direction perpendicular to the wavepacket propagation. Constructing a proper initial wavefunction is essential in the time-evolution problem. After Fourier transform, the corresponding low-energy continuum Hamiltonian near the Dirac points with a
position-dependent Dirac mass can be expressed as $h=v_{\rm F}(\tau_z\sigma_x\hat{p}_x+\sigma_y\hat{p}_y)+\sigma_zm(x,y)$, where $v_{\rm F}$ is the Fermi velocity and $\tau_z=\pm1$ label valleys K and K'. Let us focus on the simplest zero line case along $y$-axis for the specific K point, i.e., $m(x,y)=m ~{\rm sgn}(x)$ with $m$ being a constant. By solving the time-independent Dirac equation $h\Psi=E\Psi$, one obtains that when $|E|\geqslant|m|$, the eigenspinors are extended states; while when $|E|<|m|$, the eigenspinors can be written as:
\begin{eqnarray}
\begin{aligned}
\Psi_x=\exp(\frac{-|m||x|+i{\rm sgn}(m)Ey}{\hbar v_{\rm F}})\left[\begin{matrix}1-i {\rm sgn}(m)\\1+i{\rm sgn}(m)\end{matrix}\right].
\end{aligned}
\label{eq5}
\end{eqnarray}

Similarly, when the zero line is along $x$ direction, i.e., $m(x,y)=m {\rm sgn}(y)$, the eigenspinors for $|E|<|m|$ are:
\begin{eqnarray}
\begin{aligned}
\Psi_y=\exp({\frac{-|m||y|-i {\rm sgn}(m)Ex}{\hbar
v_{\rm F}}})\left[\begin{matrix}1\\-{\rm sgn}(m)\end{matrix}\right].
\end{aligned}
\label{eq6}
\end{eqnarray}

The width of wavefunction is proportional to $1/m(x,y)$. For $|E|<|m|$, based on the linear dispersion of $E=\hbar v_{\rm F} k$, Eq.~(\ref{eq6}) can be rewritten as:
\begin{eqnarray}
\begin{aligned}
\Psi_y=\exp({-\frac{|y|}{d}})\exp({-i {\rm sgn}(m)kx})\left[\begin{matrix}1\\-sgn(m)\end{matrix}\right],
%\Psi_y=\exp({-|y|/d})\exp({-i {\rm sgn}(m)kx})\left[\begin{matrix}1\\-sgn(m)\end{matrix}\right],
\end{aligned}
\label{eq7}
\end{eqnarray}
which implies that ZLM is a plane wave in the propagation direction, but becomes localized within a limited transverse direction. To construct a wavepacket at certain moment $t$, one can introduce a Gaussian term in the propagation direction. When setting the wavepacket center to be
$\vec{r_0}=(x_0,y_0)$ in real space and $k_0$ in reciprocal space,
Eq.~(\ref{eq7}) can be expressed as following:
\begin{eqnarray}
\begin{aligned}
\Psi(x,y)=
N\exp({-\frac{|x-x_0|}{d_1}-\frac{(y-y_0)^2}{2d_2^2}})
\exp({ik_0y})\left[\begin{matrix}1\\1\end{matrix}\right],
\end{aligned}
\label{eq8}
\end{eqnarray}
where $m<0$, $N$ is the normalization factor, $d_1=\hbar v_{\rm F}/|m|$, $d_2=\hbar v_{\rm F}/\Delta E$. Without loss of generality, we choose Eq.~(\ref{eq8}) as the initial wavefunction, and we take $k_0=4\sqrt3\pi/9a$ at K point [see Fig.~\ref{Fig1}(b)], with $a$ the lattice constant of graphene.

\textit{Wavepacket Dynamics at Bifurcation Point---.} Let us first examine the
wavepacket dynamics at a bifurcation point. $\langle r \rangle$ measures the
distance between wavepacket center and coordinate origin (i.e., topological
intersection), and is defined as $\langle r \rangle=\sqrt{\langle x
\rangle^2+\langle y \rangle^2}$, with $\langle x(y) \rangle=\langle\Psi| x(y)
|\Psi\rangle$. This quantity can be utilized to track the time-dependent
wavepacket trajectories. As shown in Fig.~\ref{Fig2}(a), in the straight zero
line, the wavepacket keeps moving forward along the zero line without any
backscattering, i.e., $\langle r \rangle$ versus time is almost linear. For the
right-angled zero line [see
Fig.~\ref{Fig2}(b)], the wavepacket from bottom has a probability over 99\%
turning into the left zero line, exhibiting a zero bending
resistance~\cite{vgs6} of ZLM. One can see that the minimum of $\langle r \rangle$ is nonzero, meaning that wavepacket never
reaches the exact turning point during the propagation. In Fig.~\ref{Fig2}(c), the wavepacket is equally partitioned into two parts at the bifurcation point, and $\langle r \rangle$ tends to zero when $t>30$ fs. Fig.~\ref{Fig2}(d) shows the propagation of wavepacket at a sharp turn, where the probability to the left is around 91\% and the rest turn to right. This indicates that the smaller the distance between the incoming channel and the scattering channel, the easier the scattering between the channels occurs. All these observations are exactly
consistent with the findings from electronic transport by using Green's function technique~\cite{vgs6,vgs7}, strongly suggesting the feasibility of our wavepacket dynamics in investigating the fundamental properties of ZLMs.

\textit{Unusual electronic transport at a trifurcation---.} The extremely-high-precision requirement of gating makes the AB-stacked bilayer graphene be challenging in practical application of ZLMs. Fortunately, minimally twisted bilayer graphene provides a natural system in designing ZLM-based electronics~\cite{ttb1,ttb2,ttb3,ttb4,ttb5}. However, the physical origin of current partition at the trifurcation point is still unclear, i.e., how can part of the incoming current directly pass through the trifurcation point?

To reveal the underlying physical mechanism, we apply the developed wavepacket dynamics to investigate a single node in minimally twisted bilayer graphene [see Fig.~\ref{Fig3}(a)]. For clarity, we adopt the monolayer graphene model as displayed in Fig.~\ref{Fig3}(b) with the conducting topological channels highlighted. For the ZLM encoded with certain valley K (e.g. incoming from terminal 3), the permitted outgoing terminals are terminals 2, 4 and 6. In twisted bilayer graphene, AA stacked region shrinks with decreasing twisted angle. Unless otherwise specified, in all simulations of monolayer graphene, zero-line width (corresponding to AA region in twisted bilayer graphene) is set to be the same as a single hexagonal lattice. When we set $U_1=U_2=\lambda t$, the electron wavepacket from terminal 3 is partitioned into three parts into terminals 2/4/6 at the topological intersection [see Fig.~\ref{Fig3}(c)]. Although the central metallic area is limited enough, the squared modulus of forward propagating wavepacket into terminal 6 is approximately 14\%, agreeing well with the result
from the Landauer-B$\ddot{\rm u}$ttiker formalism~\cite{ttb4}, and the rest is equally partitioned towards adjacent terminals 2 and 4. The partition to the adjacent zero lines is well understood due to the overlap between the incoming and outgoing
wavefunctions. Previously, we attributed the forward propagation to the
contribution of the narrow graphene ribbon with the same site-potential (i.e.
AA stacked ribbon with uniform bias). However, this is not true. Hereinbelow,
we will provide an analytical understanding by using the developed wavepacket
dynamics and effective models.

In Fig.~\ref{Fig2}(b), one can observe that the current never reaches the turning point when there is a angle between incoming and outgoing zero-lines. To clarify, we construct an effective model of coupled zero lines. At a circumference with a given $r$ of Fig.~\ref{Fig4}(a),
the local electronic structure of adjacent channels can be approximated by parallel topological zero lines with a distance $d$ [see Fig.~\ref{Fig4}(b)]. We attribute the interaction strength $\delta$ between zero-line modes to their mixing in the overlapping region of wave functions~\cite{ies}, i.e.,
\begin{eqnarray}
\begin{aligned}
\delta =\frac{|U|}{d}\int_{-d/2}^{d/2}\psi(x)^*_{-d/2}\psi(x)_{d/2}dx = |U|\exp[{-\frac{|U|d}{\hbar v_{\rm F}}}],
\label{eq9}
\end{aligned}
\end{eqnarray}
where
\begin{eqnarray}
\psi_{\pm d/2}=\exp [{\pm \frac{|U|}{\hbar v_{\rm F}}(x\mp d/2)}].
\label{eq10}
\end{eqnarray}

One can get the solution of $K'$ from that of K by setting $\hat{p}$ to be $-\hat{p}$. So we only need to focus on K in below. The effective
Hamiltonian of parallel zero lines at K can be written as:
\begin{eqnarray}
H(k_y)=\left[\begin{matrix}\hbar v_{\rm F} k_y & \delta\\
\delta &-\hbar v_{\rm F} k_y\end{matrix}\right],
\label{eq11}
\end{eqnarray}
where the diagonal elements describe the counter-propagating ZLMs encoded with
valley K. The energy dispersion and wavefunction are respectively:
\begin{eqnarray}
E_\pm&=&\pm \sqrt{(\hbar v_{\rm F}k_y)^2+\delta^2}, \\
\Psi_\pm&=&\frac{\exp(ik_yy)}{\sqrt{(E_\pm+\hbar v_{\rm F}
k_y)^2+\delta^2}}\left[\begin{matrix}E_\pm+\hbar v_{\rm F} k_y
\\\delta\end{matrix}\right].
\label{eq13}
\end{eqnarray}

The low-energy spectrum is displayed in Fig.~\ref{Fig4}(c). The scattering rate between adjacent channels (e.g., between states A and B at the same energy) can be given by:
\begin{eqnarray}
\begin{aligned}
T =|\langle\Psi_+(-k_y)|\Psi_+(+k_y)\rangle|^2 =(\delta/{E_+})^2,
\end{aligned}
\label{eq14}
\end{eqnarray}
which is displayed in Fig.~\ref{Fig4}(d). When the gap induced by the interaction is vanishing, it means that the electron is
located at infinity where there is no coupling between adjacent channels, thus
the scattering rate is zero. As the electron approaches the intersection,
$\delta$ increases and the electron becomes scattered to the outgoing zero
line. The Fermi energy of the electron determines its closest distance to the
intersection.

Next, we extend the effective model to six interacting counter-propagating
zero-line modes encoded with valley K at the
same circumference [see Fig.~\ref{Fig4}(a)], the corresponding Hamiltonian
is:
\begin{eqnarray}
\left[\begin{matrix}\hbar v_{\rm F} k_y & \delta &0&0&0& \delta\\
\delta &-\hbar v_{\rm F} k_y& \delta&0&0&0\\ 0&\delta&\hbar v_{\rm F}
k_y&\delta&0&0\\0&0&\delta&-\hbar v_{\rm F} k_y&\delta&0\\0&0&0&\delta&\hbar
v_{\rm F} k_y&\delta\\\delta&0&0&0&\delta&-\hbar v_{\rm F}
k_y\end{matrix}\right],
\label{eq15}
\end{eqnarray}
and the dispersion relation is:
\begin{eqnarray}
E_1\pm&=&\pm \sqrt{(\hbar v_{\rm F}k_y)^2+\delta^2}\\
E_2\pm&=&\pm \sqrt{(\hbar v_{\rm F}k_y)^2+4\delta^2},
\label{eq17}
\end{eqnarray}
where $E_1\pm$ are doubly degenerate. As the radius of the circumference decreases, the energy gap of the system becomes larger. Therefore, there is no low-energy electronic state near the trifurcation point, i.e., theoretically no electron at low energy can reach the intersection. However, why is there still forward propagation of the electronic transport?

To reveal the underlying physical origin, one can consider a scattering model as displayed in Fig.~\ref{Fig4}(a). Let us assume incoming electron (encoded with valley K) from terminal-3. It can be scattered into outgoing terminal-2 and -4. It is naturally expecting that the outgoing current at terminal-2(4) can also be scattering into incoming terminals-1(5) and -3. And subsequently, the incoming
current at terminal-1(5) can further be scattered into outgoing terminal-6 and -2(4). This indicates that the ``direct" transmitting current from terminal-3 to -6 is not so ``direct", but undergoes a complex routes along the paths of $3 \Rightarrow 2(4) \Rightarrow 1(5) \Rightarrow6$. In a recent work~\cite{ttb4}, the clue of ``bypass jump" can be inferred from the fact that forward scattering is insensitive to the increasing size of AA metallic area within a certain range.

Based on the above analysis, a topological transistor can be designed by tuning the ratio of ${U_1}/{U_2}$ to realize the on-off state [see
Fig.~\ref{Fig4}(g)], i.e., the scattering probability $T_{63}$ from terminal-3 to -6 can be continuously tuned between 0 and 1. In particular, when $U_1/U_2=0.1$, $T_{63}=0.83$ [see Fig.~\ref{Fig3}(d)]; when $U_1/U_2=4.0$, $T_{63}=0$ [see Fig.~\ref{Fig3}(e)]. In
Figs.~\ref{Fig4}(e) and \ref{Fig4}(f), one can clearly observe the direct coupling of wave functions between adjacent channels (e.g. 1 and 2) when $U_1/U_2=0.1$; on the contrary, the penetration depth of electron wavepacket in the $U_1$ regions approaches zero and the scattering by ``bypass jump" becomes vanishing when $U_1/U_2=4.0$. In Supplemental Material~\cite{smm}, we demonstrate how to manipulating $T_{63}$ via tuning the ratio of ${U_1}/{U_2}$.

\textit{Conclusion---.} In summary, a set of wavepacket dynamics and effective model of coupling ZLMs are proposed to systematically demonstrate the electronic transport properties at topological intersections. We demonstrate the zero bending resistance and the ballistic transport properties at a sharp turn from a microscopic point of view. At the topological trifurcation point in a minimally twisted bilayer graphene systems, a new current partition rule modulated by different electric gates in alternating AB and BA domains (the relations
between $U_1$ and $U_2$ in monolayer graphene) is clarified. In particular, a new scattering mechanism in the form of "bypass jump" is revealed to understand the unusual current partition at the topological trifurcation point. Our methods can not only be used to study the electronic transport properties of other topological systems, but also open up a new avenue to investigate the electronic transport behaviours of electron wavepackets in large-scale topological networks.

\begin{acknowledgements}
This work was financially supported by the NNSFC (No. 11974327), Fundamental
Research Funds for the Central Universities (WK3510000010, WK2030020032), Anhui
Initiative in Quantum Information Technologies. We also thank the
supercomputing service of AM-HPC and the Supercomputing Center of University of
Science and Technology of China for providing the high performance computing
resources.
\end{acknowledgements}

\end{document}